\documentclass[aps,prl,notitlepage,reprint]{revtex4-1}

\usepackage[colorlinks,linkcolor=red,citecolor=blue,urlcolor=red]{hyperref}
\renewcommand{\t}[1]{\mathrm{#1}}

\usepackage{graphicx} 
\usepackage{amsmath,amssymb,amsfonts}

\usepackage{amsmath,amssymb,amsfonts} % standard AMS packages
\usepackage{bm}	% bold symbols in math mode \bm{...}
\renewcommand{\mathbf}{\bm}

\usepackage{dsfont}	% proper mathbb format
\renewcommand{\mathbb}{\mathds}	% redefine \mathbb

\usepackage{mathrsfs} % use \mathscr{} for script letters in math
\usepackage{mathtools} % for proper typesetting of := and =:

\usepackage{verbatim} % use for comments
\usepackage{xcolor} % for coloring the text

\newcommand{\fref}[1]{Fig.~\ref{#1}}
\renewcommand{\eqref}[1]{Eq.~(\ref{#1})}

\begin{document}

\title{Dissipation engineering of high-stress silicon nitride nanobeams}

\author{A. H. Ghadimi, D. J. Wilson, and T. J. Kippenberg}

%\email{tobias.kippenberg@epfl.ch}

\affiliation{\'{E}cole Polytechnique F\'{e}d\'{e}rale de Lausanne (EPFL), CH-1015 Lausanne, Switzerland}
\date{\today}

\begin{abstract}
High-stress Si$_3$N$_4$ nanoresonators have become an attractive choice for electro- and optomechanical devices.  Membrane resonators can achieve quality factor ($Q$) - frequency ($f$) products exceeding $10^{13}$ Hz, enabling (in principle) quantum coherent operation at room temperature. String-like beam resonators possess conventionally 10 times smaller $Q\cdot f$ products; however, on account of their much larger $Q$-to-mass ratio and reduced mode density, they remain a canonical choice for precision force, mass, and charge sensing, and have recently enabled Heisenberg-limited position measurements at cryogenic temperatures.  Here we explore two techniques to enhance the $Q$-factor of a nanomechanical beam.  The techniques relate to two main loss mechanisms: internal loss, which dominates for large aspect ratios and $f\lesssim100$ MHz, and radiation loss, which dominates for small aspect ratios and $f\gtrsim100$ MHz. First we show that by embedding a nanobeam in a 1D phononic crystal, it is possible to localize its  flexural motion and shield it against radiation loss. Using this method, we realize $f>100$ MHz modes with $Q\sim 10^4$, consistent with internal loss and contrasting sharply with unshielded beams of similar dimensions. We then study the $Q\cdot f$ products of high-order modes of mm-long nanobeams.  Taking advantage of the mode-shape dependence of stress-induced `loss-dilution', we realize a $f\approx 4$ MHz mode with $Q\cdot f\approx9\cdot 10^{12}$ Hz.  Our results can extend room temperature quantum coherent operation to ultra-low-mass 1D nanomechanical oscillators. 
%Our results show the feasibility of ultra-low-mass 1D nanomechanical oscillators which are quantum coherent at room temperature.
\end{abstract}

\maketitle

%\begin{small}

Silicon nitride (SiN) nanomembranes and beams have emerged as striking exceptions to the empirical $Q\propto V^{1/3}$ scaling of solid-state mechanical oscillators \cite{imboden2014dissipation}.  Much of the recent history of 
this development can be traced to the `discovery' that a commercial TEM slide made of 100 nm-thick LPCVD Si$_3$N$_4$, forming a membrane resonator,
can realize a room temperature $Q\cdot f>6\cdot 10^{12}$ Hz as well as ppm-level optical loss \cite{zwickl_high_2008,wilson2009cavity}.  The former constitutes a basic requirement for quantum-coherent evolution (since in this case the oscillator's frequency $f$ is larger than its thermal decoherence rate, $k_{B}T/hQ$), while the latter enables integration into a high finesse optical cavity, allowing for quantum-limited displacement readout and actuation \cite{thompson_strong_2008,purdy_observation_2013}. Encouraged by these prospects, considerable effort has been made to understand the source of high $Q$ in SiN films. It is now generally accepted to be the large ($\sim 1$ GPa) tensile stress resulting from LPCVD, which increases the elastic energy stored in the film without changing the material loss tangent \cite{unterreithmeier2010damping,schmid2011damping,yu2012control,villanueva2014evidence} (a concept in fact dating back to mirror pendulum supports in gravitational wave interferometers \cite{gonzalez1994brownian}).
Exploiting this insight, radio frequency membranes with room temperature $Q\cdot f>10^{13}$ Hz have recently been realized, using a combination of high order modes \cite{wilson2009cavity,chakram2014dissipation}, stress engineering \cite{norte2015mechanical}, and acoustically-shielded supports \cite{tsaturyan2014demonstration,yu2014phononic}.

\begin{figure}[b!]
	\includegraphics[width=0.95\columnwidth]{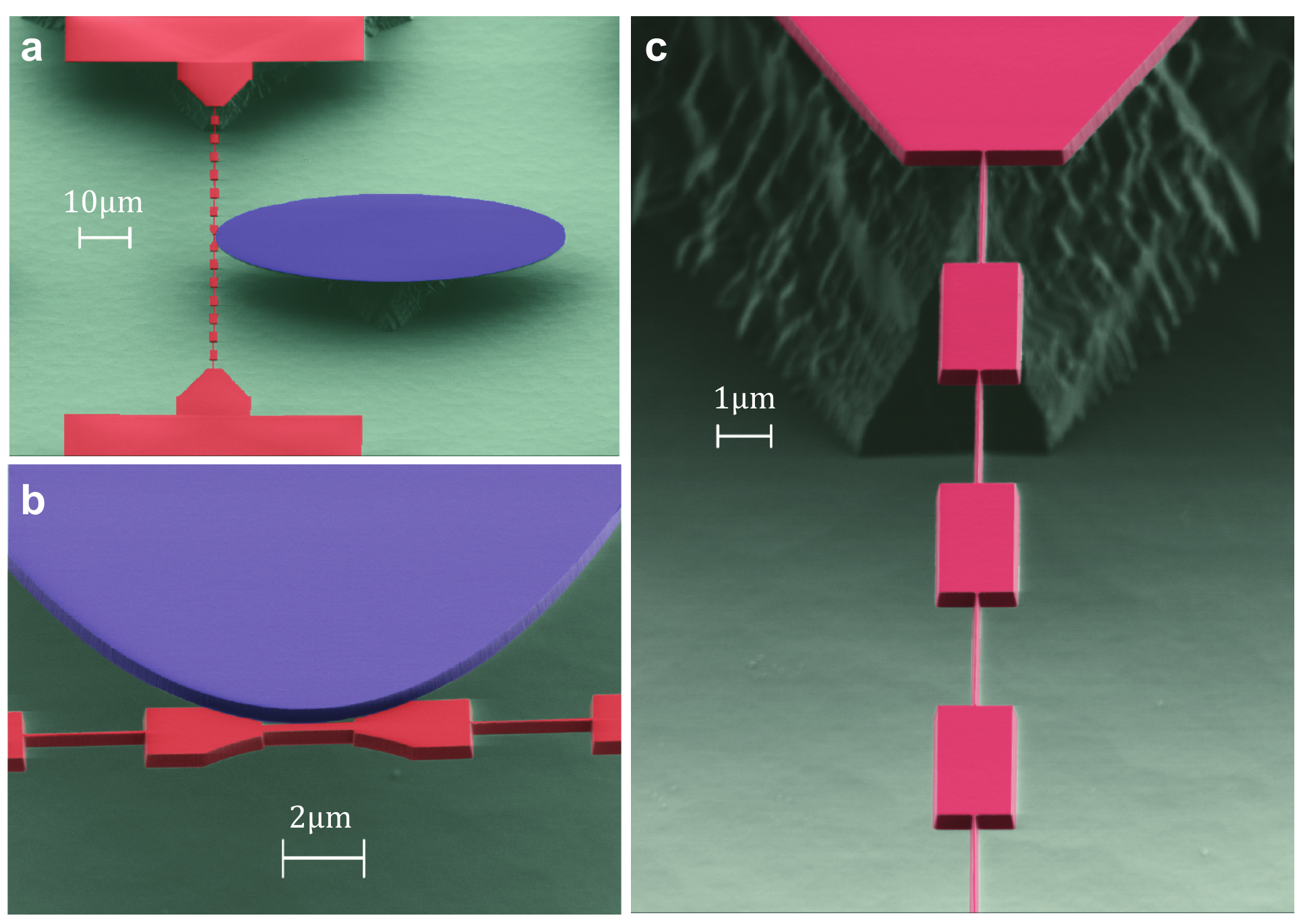}
	\caption{\textbf{A nanobeam embedded in a 1D phononic crystal}. SEM images of the device. False-coloring is used to distinguish the Si substrate (green), the micro-patterned SiN nanobeam (red) and a SiN microdisk (blue).  The microdisk is used to optically probe the beam's displacement \cite{anetsberger_near-field_2009}.   }
	\label{fig:SEM}
\end{figure}

\begin{figure*}[t!]
	\centering
	\includegraphics[width=0.9\linewidth]{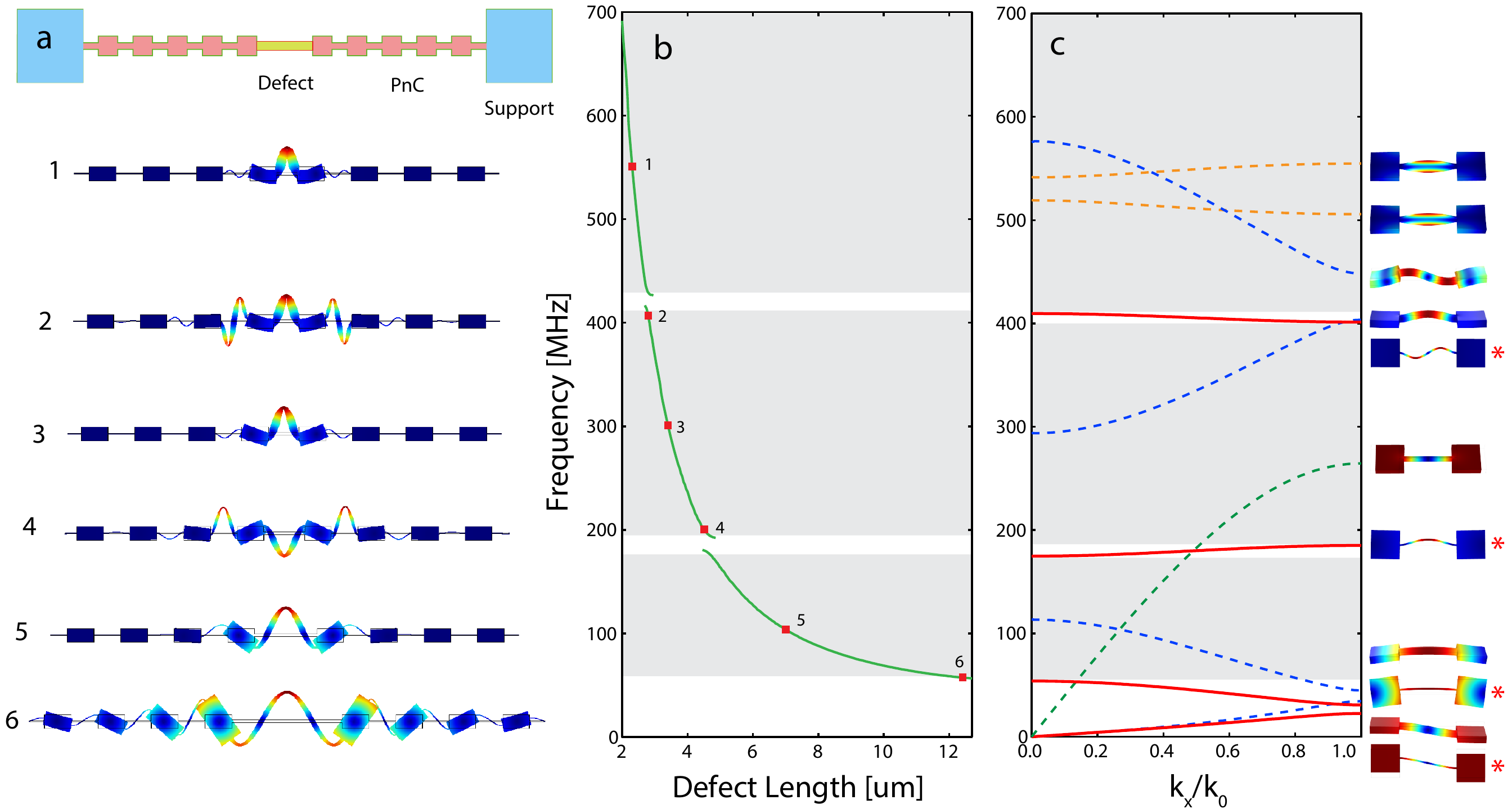}
	\caption{\textbf{Simulation of defect modes and the phononic crystal bandgap.} (a) Schematic diagram of the device (top view), consisting of a nanobeam (green) patterned as a defect in a 1D phononic crystal (pink). Unit cell dimensions used in this work are ($h_1$,$L_1$)=(100 nm,3 $\mu$m) for the short segment and ($h_2$,$L_2$)=(1.5 $\mu$m,3 $\mu$m) for the long segment. The SiN thickness of all devices is 400  nm. (b) Simulated fundamental eigenfrequency of a defect beam with in-plane thickness $h=100$ nm and length $L=2-16\,\mu$m. Gray regions indicate bandgaps for in-plane flexural modes.  Six representative modes are highlighted; their FEM-simulated modeshapes are plotted on the left.  (c) Dispersion diagram of the phononic crystal. Curves represent different modes of the unit cell, color-coded according to symmetry (red, blue, green, and orange correspond to in-plane, out-of-plane, breathing, and torsional modes, respectively).  Gray regions correspond to pseudo-bandgaps for in-plane modes.}
	\label{fig:PhonoicCrystal}
	%\vspace{-3pt}
\end{figure*}

In this report we study two strategies for reducing the dissipation of a high-stress SiN nanobeam, the 1D analog of a membrane which is in principle capable of $Q\cdot f>10^{13}$ Hz with orders of magnitude lower mass and significantly reduced mode density, making it suitable for precision mass \cite{jensen2008atomic}, charge \cite{cleland1998nanometre}, and force \cite{mamin2001sub} sensing, in addition to quantum optomechanics experiments \cite{anetsberger_near-field_2009,wilson2015measurement}.  The major challenge associated with nanobeams is their small form factor, which makes it difficult to transduce their motion.  Towards this end, we employ a microcavity-based near-field sensor \cite{anetsberger_near-field_2009} capable of non-invasive thermal noise measurements with $\t{fm}/\sqrt{\t{Hz}}$ resolution.  Our first strategy is to pattern short ($<10\,\mu$m), very-high-frequency ($>100$ MHz) nanobeams \cite{huang2005vhf}, for which an important source of loss is acoustic radiation.  To address this challenge, we pattern the nanobeam as a defect in a 1D phononic crystal (PnC). When the frequency of a beam mode coincides with a band-gap of the PnC, the latter acts as an acoustic shield, effectively localizing flexural motion and suppressing radiation loss \cite{goettler2010realizing,mohammadi2011simultaneous,hsu2011design}.  Using this strategy, we realize localized flexural modes with effective mass $m< 1$ pg,  $f>100$ MHz and $Q\sim10^4$, consistent with internal loss and contrasting sharply with unshielded beams of similar dimensions. Our second strategy is to study the high order modes of mm-long nanobeams, which exhibit enhanced $Q\cdot f$ due to tensile stress. We demonstrate that by appropriate choice of beam length and mode order, it is possible to realize $m\sim 10\,\t{pg}$, $f\sim10$ MHz modes with a `quantum-enabled' $Q\cdot f\approx 10^{13}$  Hz, a regime previously accessed only by $m\sim 10$ ng membrane resonators \cite{wilson2009cavity,chakram2014dissipation,tsaturyan2014demonstration}.

We first consider the device shown in \fref{fig:SEM}.  Here a high-stress Si$_3$N$_4$ thin film has been patterned into a 1D PnC with a beam-like defect at its center.  From the standpoint of the beam, the crystal acts like a radiation shield \cite{yu2014phononic,tsaturyan2014demonstration,chan2012optimized}, the performance of which is determined by the band structure of the box-shaped unit cell. A simulation of the dispersion diagram of the unit cell is shown in \fref{fig:PhonoicCrystal}c, with lines of different color corresponding to modes of the cell with different symmetries. In this work, we consider only in-plane flexural modes because of their compatibility with displacement readout (using the microcavity-based sensor shown in blue in \fref{fig:SEM}). Pseudo-bandgaps for in-plane symmetry are indicated by gray shading in \fref{fig:PhonoicCrystal}c.  The presence of a bandgap implies strong reflection of waves from the PnC. It also implies the support of localized defect modes. To illustrate this concept, a simulation of the fundamental flexural mode of a beam embedded in a 14-element PnC is shown in \fref{fig:PhonoicCrystal}b.  Dimensions of the beam and unit cell are given in the caption. The beam's length is varied to span the frequency range of the three bandgaps shown in \fref{fig:PhonoicCrystal}c.  Qualitatively, it is evident that modes which are well-centered in a bandgap (1,3,5) exhibit strong confinement.  Conversely, modes near the edge of a bandgap (2,4,6) penetrate deeply into the PnC. This is because the defect modes start to hybridize with the in-plane modes of the unit cell, shown as red curves in \fref{fig:PhonoicCrystal}c. 

Our central claim is that localized defect modes will exhibit reduced radiation loss and similar effective mass relative to unshielded beams. To demonstrate this concept, shielded and unshielded beams of various lengths were fabricated (see \cite{suppinfo}) and tested.  To probe mechanical displacement, an elliptical microdisk cavity is patterned next to each beam, separated by $\sim 80$ nm.  Whispering gallery modes of the microdisk are excited with a detuned 1550 nm laser field using a tapered optical fiber, enabling evanescent displacement readout \cite{anetsberger_near-field_2009} with an imprecision of  $\sim$ 1 fm$/\sqrt\text{Hz}$. All measurements were performed in a vacuum chamber at $10^{-4}$ mbar in order to reduce gas damping.  In conjunction with the relatively large cavity linewidth of $\sim1$ GHz (mitigating radiation pressure effects), this enables non-invasive thermal noise measurements for beams as short as 4 $\mu$m.

\begin{figure}[t!]
	\centering
	\includegraphics[width=1\linewidth]{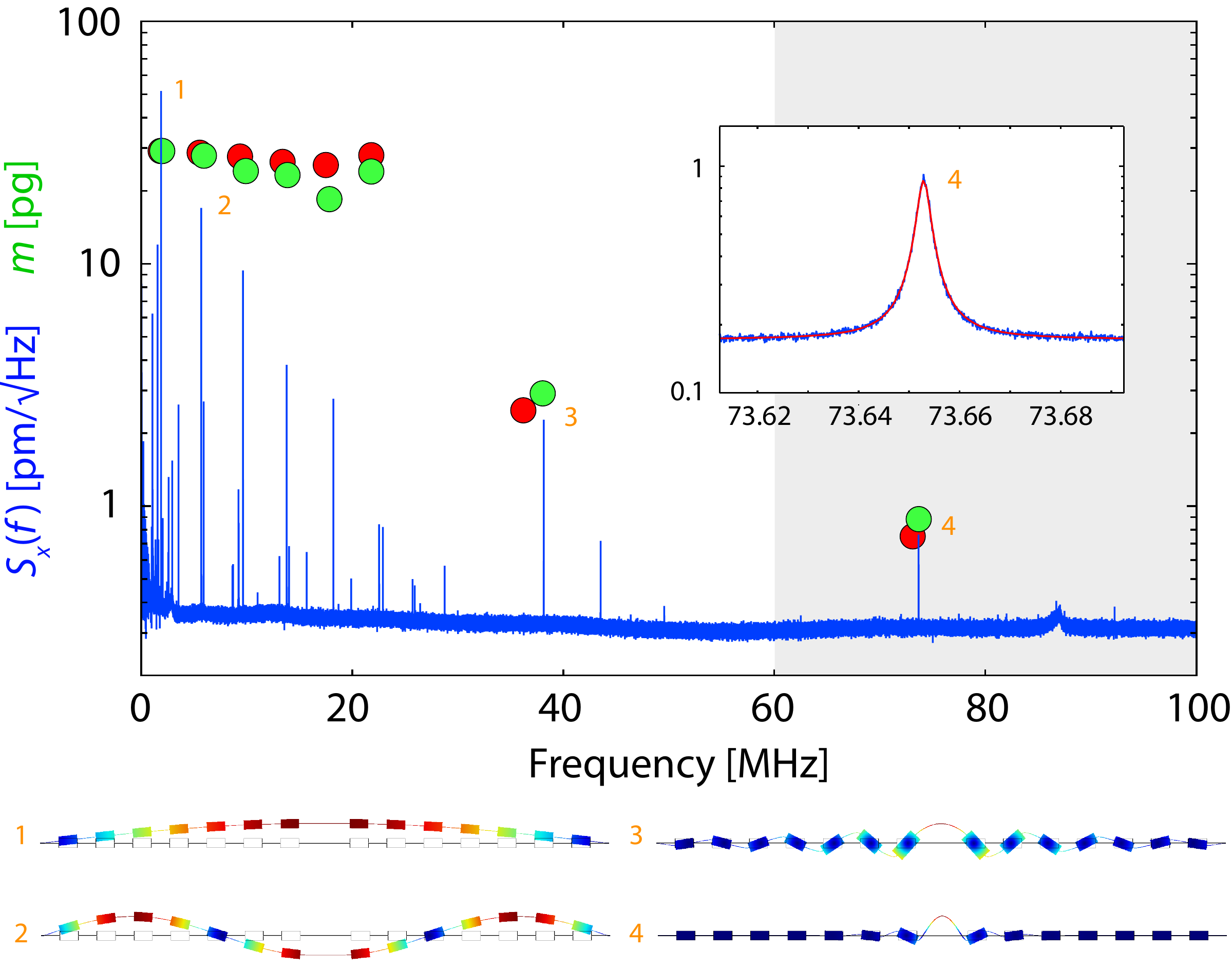}
	\caption{\textbf{Broadband displacement spectrum of a PnC-shielded nanobeam}. A beam length of 8.25 $\mu$m is shown. FEM-simulations of four representative mode shapes are plotted below. Green points above selected peaks indicate the inferred effective mass. Red points are obtained from an FEM model.  The absolute magnitude of the displacement spectrum and the effective mass is estimated by bootstrapping the latter to an FEM model (red points) of the effective mass for mode $\#$1. Shaded regions correspond to the PnC bandgaps in \fref{fig:PhonoicCrystal}. Inset: Magnified displacement spectrum of the localized (defect) mode, fitted to a Lorentzian. }
	\label{fig:Spectrum}
	\vspace{-3mm}
\end{figure}

As a demonstration of spatial mode confinement, the thermal displacement noise spectrum of a 5 $\mu$m-long defect embedded in a 100 $\mu$m-long, 14-cell PnC is presented in \fref{fig:Spectrum}.  The fundamental in-plane mode of the defect appears at 74 MHz, situated within a large, spectrally quiet window coinciding with the pseudo-bandgap of the PnC.  The small effective mass of the defect mode manifests in the relatively large area beneath the thermal noise peak $\langle x^2\rangle \approx k_\t{B}T/(4m\pi^2 f^2)$.  Comparing $(\langle x^2\rangle f^2)^{-1}$ for the defect mode to that of adjacent peaks (green points) reveals a 1000-fold decrease in $m$ relative to the fundamental in-plane mode of the extended structure.  Quantitative agreement of this scaling with an FEM simulation (red points, assuming a point-like probed at the midpoint of the defect) corroborates an estimated effective mass of $m\approx 1$ pg for the localized mode.  The corresponding fitted mechanical linewidth is $\gamma = 5.1$ kHz ($Q \equiv f/\gamma = 1.4\cdot 10^4$), giving access to a low thermal-noise-limited force sensitivity of $8\pi k_{B}Tm\gamma=(0.7\,\t{fN}/\sqrt{\t{Hz}})^2$.

\begin{figure}[b!]
	\vspace{-3mm}
	\centering
	\includegraphics[width=1\linewidth]{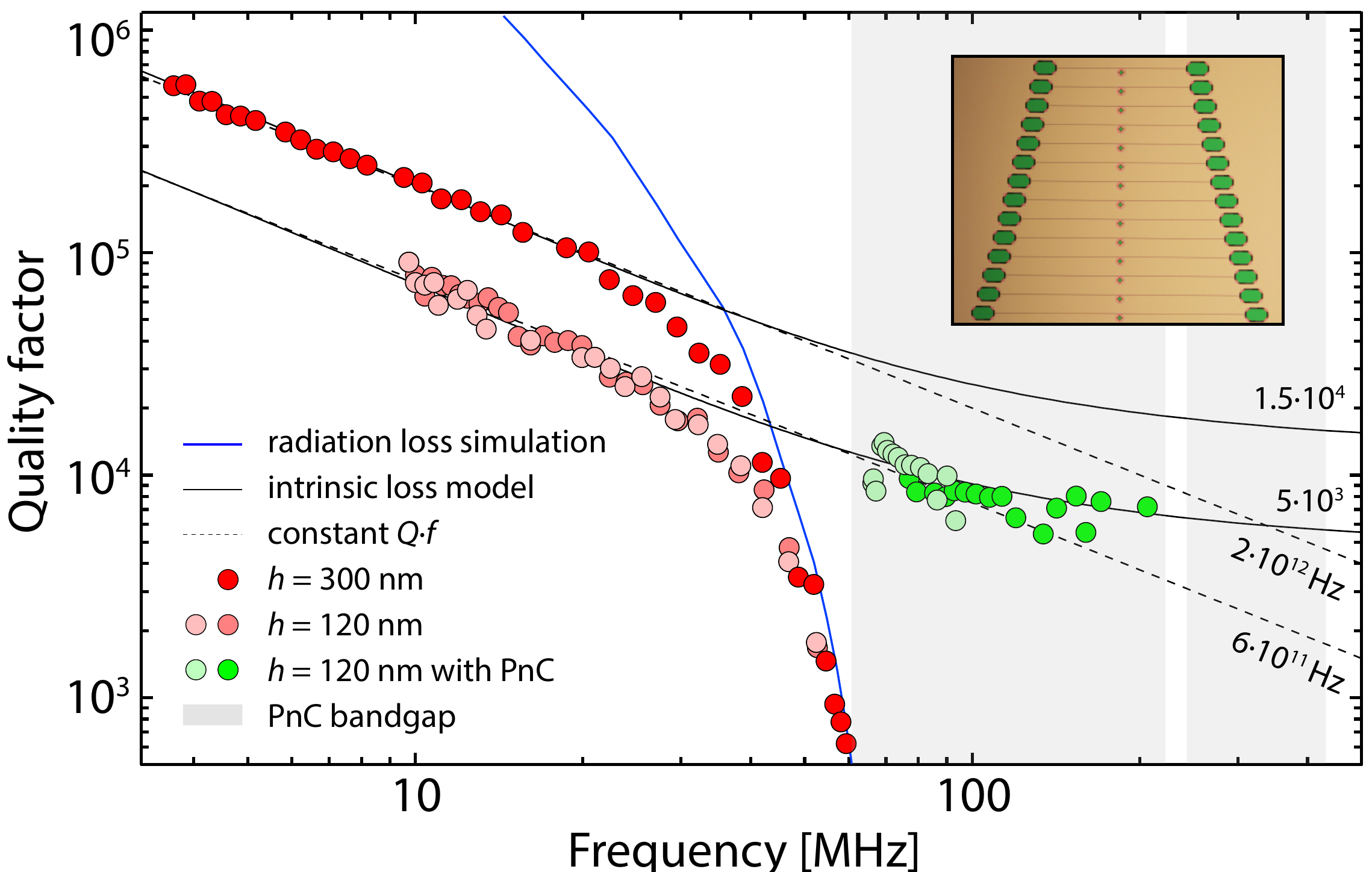}
	\caption{\textbf{Quality factor of nanobeams with and without PnC shield}. $Q$ versus $f$ for the fundamental mode of nanobeams of various lengths ($L\approx 4-90\,\mu$m).  Green (red) points correspond to beams with (without) a PnC shield. Solid black lines are a single parameter fit to the internal loss model in \eqref{eq:Qinternal}, using $Q_\text{int}$ as the fit parameter. Dashed black lines indicate constant $Q\cdot f$. The solid blue line is an FEM simulation of radiation loss for the unshielded beams.  Shaded regions correspond to the PnC bandgaps in \fref{fig:PhonoicCrystal}. The inset shows a sample chip containing a typical set of devices with different beam length, in this case without a PnC. }
	\label{fig:sweep_result}
	\vspace{-0mm}
\end{figure}

To assess the performance of the PnC as an acoustic shield, we have made a comprehensive study of $Q$ vs $f$ for shielded and unshielded beams of various lengths, using thermal noise measurements as in \fref{fig:Spectrum}.  Results are compiled in \fref{fig:sweep_result}.  Green and red points correspond to shielded and unshielded beams with lengths of $L = 12.2 - 4.1\,\mu$m and $10.5 - 90.5\,\mu$m, respectively, comprising a total of 121 independent devices. The most striking feature is a sharp transition at $f_0\sim 50$ MHz, corresponding to $L\sim10\,\mu$m, at which the $Q$ of unshielded beams changes from $Q\propto f^{-1}$ to a $Q\propto e^{-\alpha(f-f_0)}$.  By contrast, shielded beams with $L<10\,\mu$m exhibit a roughly constant $Q$, suggesting a qualitatively different loss mechanism compared to unshielded beams.

To understand the scaling of $Q$ in \fref{fig:sweep_result}, we consider a model of the form $Q^{-1} = Q_\t{int}^{-1}+Q_\t{rad}^{-1}$, where $Q_\t{int}$ and $Q_\t{rad}$ are models of loss due to internal friction and radiation of acoustic waves (phonon tunneling \cite{wilson2008intrinsic}) into the solid-state supports, respectively.  We first consider internal loss, adopting a standard model that treats the film as an anelastic plate subject to cyclic loading \cite{gonzalez1994brownian,yu2012control,schmid2011damping,villanueva2014evidence,unterreithmeier2010damping}.  In this approach, friction is treated as a delay between internal strain and stress (i.e., a complex Young's modulus, $E$) \cite{zener1938internal}.  The enhanced $Q$ of stressed films is related to a renormalization of the total elastic energy, viz., adding tensile stress $\sigma$ contributes an energy due to elongation, $U_\t{elong}\propto\sigma$, without substantially increasing the energy stored in bending, $U_\t{bend}\propto E$ (which gives rise to dissipation) \cite{unterreithmeier2010damping}.  The $Q$ of a stressed film is thus enhanced relative to the unstressed case $Q_{0}=\t{Re}[E]/\t{Im}[E]$ by approximately $Q/Q_{0}\approx 1+U_\t{elong}/U_\t{bend}$.  The magnitude of the `loss-dilution' factor $Q/Q_{0}$ depends on the stressed mode shape. For the string-like modes of stressed beams with thickness $h\ll L$, it can be shown that \cite{yu2012control,villanueva2014evidence,unterreithmeier2010damping} 
\begin{equation}\label{eq:Qinternal}
	Q_\t{int}(\sigma,n)\approx Q_0\left(1+\left(\underbrace{2\lambda}_\t{clamping}+\underbrace{n^2\pi^2\lambda^2}_\t{antinode}\right)^{-1}\right)
\end{equation}
where $\lambda = (h/L)\sqrt{E/12\sigma}$.  Underbrackets here indicate contributions due to bending at the clamping points and at the antinodes of the flexural mode, respectively, which depends on the mode order, $n$.  The value of $\lambda$ can be estimated from the dispersion relation
\begin{equation}\label{eq:fdispersion}
f(\sigma,n)\approx f_0 n \sqrt{\left(1+n^2\pi^2\lambda^2\right)},
\end{equation}
where $f_{0}\approx (1/2L)\sqrt{\sigma/\rho}$.

\begin{figure}[t]
	\centering
	\includegraphics[width=1\linewidth]{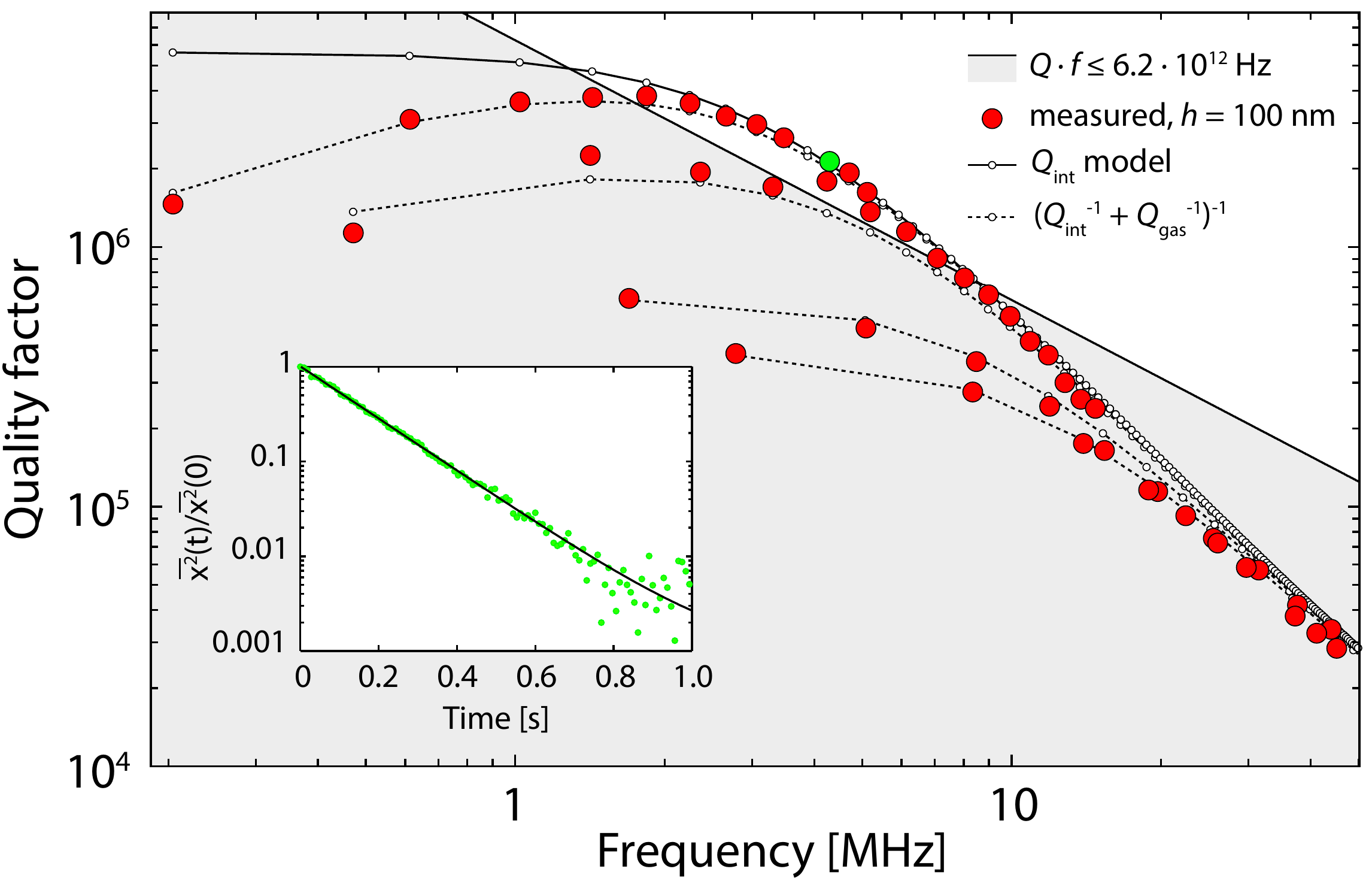}
	\caption{\textbf{Quality factor versus mode order for long high-stress SiN nanobeams}. $Q$ versus frequency for the odd-ordered flexural modes of nanobeams with lengths $L=1.28,\,0.55,\,0.15,\,\t{and}\,0.095$ mm. Gray shading indicates $Q\cdot f < k_B\cdot 300\,\t{K}/h=6.2\cdot 10^{12}$ Hz. Black lines correspond to models for $Q$ including internal loss (\eqref{eq:Qinternal}) and an estimated gas damping rate of $\gamma_\t{gas}=0.15\,\t{Hz}$ ($Q_\t{gas}=  f/(0.15\,\t{Hz})$), using $\lambda = 5.5\cdot h/L$. Inset: Ringdown measurement used to obtain the green point, $Q\cdot f\approx 9.0\cdot 10^{12}$.}
	\label{fig:long_beam}
	\vspace{-3mm}
\end{figure}

\eqref{eq:Qinternal} and \eqref{eq:fdispersion} are used to model the internal loss of beams with two widths ($h=120,300$ nm) in \fref{fig:sweep_result} (solid black lines).  A value of $\lambda\approx 5\cdot (h/L)$ is obtained by fitting measurements of $f$ versus $n$ to \eqref{eq:fdispersion} (see \fref{fig:long_beam}).  A value of $Q_0\approx 5\cdot10^3\cdot(h/100\,\t{nm})$ is obtained by bootstrapping to the $Q\propto f^{-1}$ scaling of long ($\lambda\ll1$) unshielded beams.  The inferred surface loss coefficient of $Q_0/h\approx 50\,\t{nm}^{-1}$ agrees well with the findings of Villanueva \emph{et. al.} \cite{villanueva2014evidence}; however, $Q_0$ cannot be accessed directly with unshielded beams due the large deviation from the internal loss model for $f\gtrsim 50$ MHz $(L\lesssim 10\,\mu\t{m}$).  Remarkably, the $Q$ of shielded beams (green points) recovers to the expected internal loss scaling for $f\sim 50-200$ MHz  $(L\sim 4-12\,\mu\t{m}$), suggesting that the loss mechanism is acoustic radiation.

To confirm that the deviation from the internal loss model in \fref{fig:sweep_result} is due to acoustic radiation, we have conducted a no-free-parameter finite element simulation using COMSOL (for details see \cite{suppinfo}). The geometry and stress-profile of the beam and PnC are determined from SEM imaging. The result of the simulation is shown as a solid blue line \fref{fig:sweep_result} and agrees qualitatively well with the data.  Notably, the  cutoff frequency for radiation losses depends sensitively son the geometry of the beam near the end-supports, as well as the support pillars (also included in the model).  Chamfering the beam at its ends, as shown in \fref{fig:SEM} reduces the knee to $f\sim50$ MHz, which is lower than has been observed for beams of similar dimensions with no end-chamfer \cite{huang2005vhf}.

Having addressed the challenge of reducing radiation loss in short, VHF beams, we now study a strategy for achieving high $Q\cdot f$ while keeping $m$ fixed, by using higher order modes. Assuming $\lambda\ll1$ and treating $n$ as a continuous variable in \eqref{eq:Qinternal} and \eqref{eq:fdispersion}, the maximum $Q\cdot f$ is obtained at $n_\t{opt}\approx\sqrt{2/(\pi^2\lambda)}$ and is given by
\begin{equation}\label{eq:Qfmax}\begin{split}
\frac{Q_0f_0}{\sqrt{8\pi^2\lambda^3}}&\approx 1.2\cdot 10^{13}\,\t{Hz}\cdot\frac{Q_0/h}{50\,\t{nm}^{-1}}\sqrt{\frac{L}{1\,\t{mm}}\frac{100\,\t{nm}}{h}}\\ &\cdot\left(\frac{\sigma}{0.8\,\t{GPa}}\right)^{\tfrac{5}{4}}\left(\frac{300\,\t{GPa}}{E}\right)^{\tfrac{3}{4}}\sqrt{\frac{2.7\,\tfrac{\t{kg}}{\t{cm}^3}}{\rho}}.
\end{split}\end{equation}
%\begin{equation}
%\frac{Q_0 f_0}{\sqrt{8\pi^2\lambda^3}}\approx Q_0\cdot\frac{L^{1/2}\cdot\sigma^{5/4}}{h^{3/2}\cdot E^{3/4}\cdot\rho^{1/2}}}\cdot\frac{3^{3/4}}{2\pi}
%\end{equation}
%\begin{equation}
%Q_\t{int}(\sigma,n_\t{opt})\cdot f(\sigma,n_\t{opt})\approx Q_0 f_0\sqrt{\frac{1}{8\pi^2\lambda^3}}
%\end{equation}
Therefore high $Q\cdot f$ may be obtained by using high order modes of long and thin beams. We note that demonstrations of high-stress Si$_3$N$_4$  membranes with $Q\cdot f >10^{13}$ Hz also exploit this scaling, with the mode index $n$ in \eqref{eq:Qinternal} and \eqref{eq:fdispersion} replaced by $n^2\rightarrow n^2+m^2$.  For diagonal modes ($n=m$), this results in an increase of \eqref{eq:Qfmax} by a factor of $\sqrt{2}$.  Interestingly, the conventionally higher $Q\cdot f$ reported for membranes as compared to nanobeams appears to be related to the fact that larger membranes ($L\sim 1 $ mm) have been studied \cite{wilson2009cavity,yu2012control,chakram2014dissipation}.

To explore the prediction made in \eqref{eq:Qfmax}, we have measured $Q(n)$ and $f(n)$ for odd-ordered modes of unshielded beams with thickness $h=100$ nm and lengths varying from $L=0.95-1.28$ mm.  Results are shown in \ref{fig:long_beam}. To mitigate systematic error associated with the sub-Hz mechanical linewidths, in this case $Q$ was extracted from ringdown measurements.  To perform a ringdown, the beam is resonantly excited using a piezo located beneath the sample chip; the drive is then shuttered off while displacement is continuously recorded using a network analyzer (with bandwidth $B\gg\gamma$). For the longest beam, $L = 1.28\;\t{mm}$, we observe a maximum of $Q\cdot f\approx 9.0\cdot10^{12}$ for $n=21\,(f\approx 4\,\t{MHz})$.  We compare this to a model curve for $Q(n)\cdot f(n)$ (solid black line) based on a value of $\lambda=5\cdot h/L$ estimated by fitting the dispersion relation $f(n)$ (\eqref{eq:fdispersion}).  The model predicts $Q(n_\t{opt})\cdot f(n_\t{opt})$ to within $10\%$, but qualitatively overestimates $Q$ for lower frequencies.  This discrepancy is due to gas damping, which contributes a systematic additional damping of $\gamma_\t{gas}\approx0.15$ Hz.  

We emphasize that while the concept of PnC acoustic shielding has recently been explored in the context of nanomembranes \cite{yu2014phononic,tsaturyan2014demonstration} and nanobeams \cite{chan2012optimized}, our results, to the best of the authors' knowledge, represent the first explicit demonstration that the intrinsic $Q$ of a nanomechanical resonator can be recovered in the presence of otherwise dominant radiation loss. Moreover, the value of $Q\cdot f\approx 9.0\cdot  10^{12}$ demonstrated in \fref{fig:long_beam} establishes a new benchmark for high-stress SiN nanobeams.  In conjunction with the large zero-point motion and sparse mode density of nanobeams, %these complementary dissipation engineering strategies may enable a new class of quantum electro- and optomechanical devices based on Heisenberg-limited measurements \cite{wilson2015measurement}, perhaps even at room temperature \cite{norte2015mechanical,schilling2016near}. 
and building on recent advances in near-field cavity optomechanical coupling \cite{wilson2015measurement,schilling2016near}, we envision application of these dissipation engineering strategies towards a new generation of quantum optomechanics experiments based on Heisenberg-limited measurements \cite{wilson2015measurement}, perhaps even at room temperature. 
Nanoscale PnCs are furthermore readily extended to alternative materials such as diamond \cite{khanaliloo2015single}, %and may enable new investigations into the microphysics of internal loss \cite{groblacher_observation_2015}. 
opening the door for studying the microphysics of internal loss \cite{groblacher_observation_2015}.  
PnC-shielded nanoresonators may also find application as high frequency mechanical filters and oscillators \cite{teva2008vhf,feng2008self}.

%\begin{center}
%	\textbf{ACKNOWLEDGMENTS}
%\end{center}

The authors thank Dr. Michael Zervas and Dr. Zdenek Benes for fabrication advice. All samples were fabricated at the CMi (Center of MicroNanotechnology) at EPFL. Research was carried out with support from an ERC Advanced Grant (QuREM), from the Swiss National Science Foundation and through grants from the NCCR of Quantum Engineering (QSIT). D.J.W. acknowledges support from the European Commission through a Marie Skłodowska-Curie Fellowship (IIF project 331985).

%\end{small}
%\bibliographystyle{apsrev4-1}

%\input{ref1.bbl}
%\bibliography{Amir_ref,ref1}
%\bibliography{ref1}

%\clearpage

\onecolumngrid
\clearpage

\begin{center}
	\large{\textbf{Supplementary information for\\ 
			``Dissipation engineering of high-stress silicon nitride nanobeams''}}
\end{center}
	
\setcounter{equation}{0}
\setcounter{figure}{0}
\setcounter{table}{0}
	
	%%%%%%%%%%%%%%%%%%%%%%%%%%%%%%%%%%% BEGIN MATTER %%%%%%%%%%%%%%%%%%%%%%%%%%%%%%%%%%%%
	
	%\tableofcontents
	%\addtocontents{toc}{\protect\setcounter{tocdepth}{1}}
	
	\section{I.\hspace{5mm}Sample fabrication}\label{sec:fab}
	
	\begin{figure*}[b!]
		\includegraphics[scale=0.7]{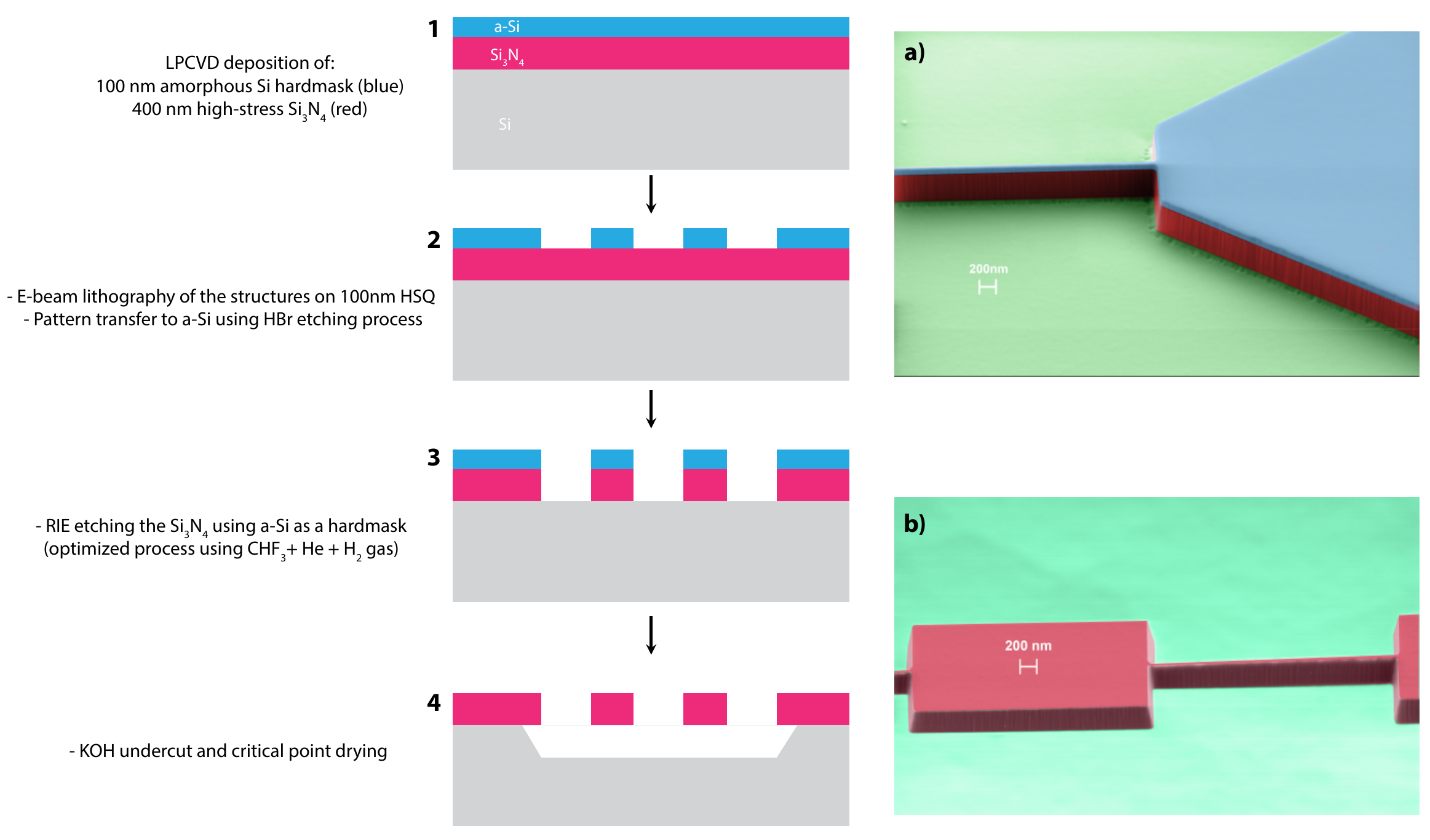}
		\caption{\textbf{Fabrication process flow.} (1-4) Illustration of the main fabrication steps.  (a) False-colored SEM images of the device after step 3, with green, blue, and red indicating Si, a-Si, and Si$_3$N$_4$, respectively. (b) False-colored SEM image of a single PnC unit cell after KOH undercut (step 4).}
		\label{fig:process_flow}
	\end{figure*}
	
	\subsection{A.\hspace{5mm}Overview}
	
	Nanobeam-microdisk samples as shown in Fig. 1 of the main text are fabricated according to the procedure outlined in Fig. 1.  The process begins with a series of low pressure chemical vapor deposition (LPCVD) steps on a 4" CZ Si wafer.  400 nm of stoichiometric Si$_3$N$_4$ (SiN) is deposited at 800$^\circ$C, followed by 100 nm of amorphous Si (a-Si) at 500$^\circ$C (to be used as a hard mask).  %After cool-down, thermal expansion mismatch between the SiN film and the Si substrate results in a tensile stress of approximately $0.8$ GPa in the SiN film. 
	The microdisk and nanobeam are patterned on top of the a-Si film by electron beam lithography, in 100 nm of HSQ. The pattern is transfered into the a-Si film (creating a hard mask for the subsequent SiN etch) using a HBr etch process.  The pattern is then transfered into the SiN film by reactive ion etching, using an SPTS APS Dielectric Etcher and CHF$_3$ chemistry. The process is concluded by undercutting the microdisk and nanobeam with KOH, followed by critical point drying.  
	
	\subsection{B.\hspace{5mm}Fabrication considerations related optomechanical coupling}
	The fabrication process was developed and optimized with two goals in mind: (1) to minimize the in-plane gap ($x$) between the nanobeam and the microdisk and (2) to minimize the sidewall roughness of the microdisk.  Small $x$ is desirable as it gives rise to a large parametric (optomechanical) coupling $G=\partial\omega_\t{c}/\partial x$ between the lateral position of the beam and the frequency $\omega_\t{c}$ of the microdisk whispering gallery mode (WGM) \cite{aspelmeyer_cavity_2015}.  For beams with in-plane thickness much smaller than the operating wavelength $\lambda$, it can be shown that $G\propto e^{-x/x_\t{ev}}$, where $x_\t{ev}\sim\lambda/10$ is the in-plane evanescent decay length of the WGM \cite{anetsberger_near-field_2009}.  Thus a design gap of $x<100$ nm was chosen in conjunction with an operating wavelength of $\lambda\approx 1550$ nm. Low sidewall roughness is desirable in order to reduce the possibility of WGM loss ($\kappa$) due to Rayleigh scattering from the disk's surface \cite{borselli2004rayleigh}.  Collectively, increasing $G$ and reducing $\kappa$ was crucial to achieving the high sensitivity displacement measurements shown in Fig. 3 and employed in Figs. 4-5 of the main text.  For example, in Fig. 3 of the main text, the thermomechanical signal-to-noise (in this case dominated by electronic detector noise) is proportional to $(G^2 P/\kappa)^2$, where $P$ is the optical input power \cite{aspelmeyer_cavity_2015}.  
	
	The requirements for etching optically smooth sidewalls separated by $x<100$ nm depends on the thickness of the SiN film. We chose a $t=400$ nm-thick film as a compromise between achieving a large WGM evanescence and low WGM radiation loss.  Achieving a vertical $x<100$ nm gap thus required etching SiN with an aspect ratio much better than 1:4.  We found that achieving this aspect ratio while maintaining an optically smooth surface was difficult with standard RIE etching techniques (employing standard e-beam resists), due to mask erosion.  It is for this reason that a-Si was employed as a secondary hard mask. a-Si provides high selectivity ($>$1:10) relative to SiN in RIE as well as high resistance to mask erosion. Thus we have achieved gaps as small as $x\approx 80$ nm and relatively smooth sidewalls, characterized by optical quality factors in excess of $10^5$ in the absence of a beam (see Fig. 2).  For the results shown in Figs. 3-5 of the main text, we employed elliptical microdisks with a fixed circumference of approximately $30\,\mu$m, an ellipticity (major radius over minor radius) ranging from 1 to 4, and $x = 80-200$ nm.  Typical intrinsic optical linewidths of $\kappa/2\pi = 1-10$ GHz were observed.  An estimate of $G(x=100\,\t{nm})\sim100$ MHz/nm is obtained by approximating the microdisk WGM profile with the analytical form of a microtoroid WGM, as discussed in \cite{anetsberger_near-field_2009,schilling2016near}.
	
	\begin{figure*}[t]
		\centering
		\includegraphics[width=0.6\linewidth]{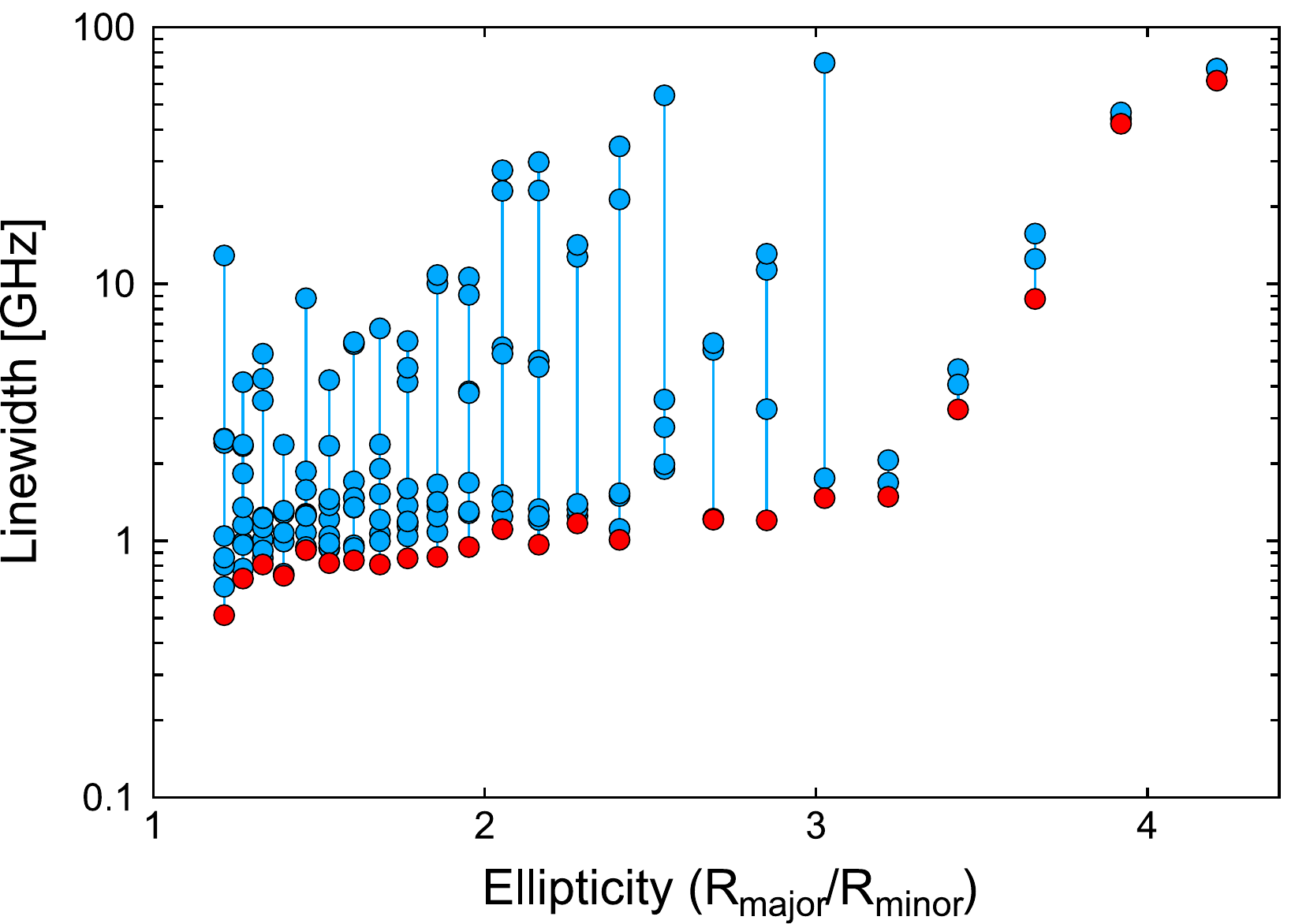}
		\caption{\textbf{Optical loss of elliptical microdisk cavities.} Measured WGM linewidth for microdisks with a fixed circumference of 31 $\mu$m and ellipticity (ratio of the major and minor radii) varying from 1 to 4.  WGMs of different polarization and radial order were measured, giving rise observed variation in optical loss for each ellipticity. Red points highlight the best value from each measurement set.}
		\label{fig:radiation_FEM}
		%\vspace{-3pt}
	\end{figure*}

	\section{II.\hspace{5mm}Acoustic radiation loss model}\label{sec:sim}
	
	% To correct simulate the  a stress relaxation simulation is necessary  (to calculate the stress profile) before solving for Eigenmodes. 
	
	\subsection{A.\hspace{5mm}Overview}
	
	\begin{figure*}[t]
		\centering
		\includegraphics[width=0.95\linewidth]{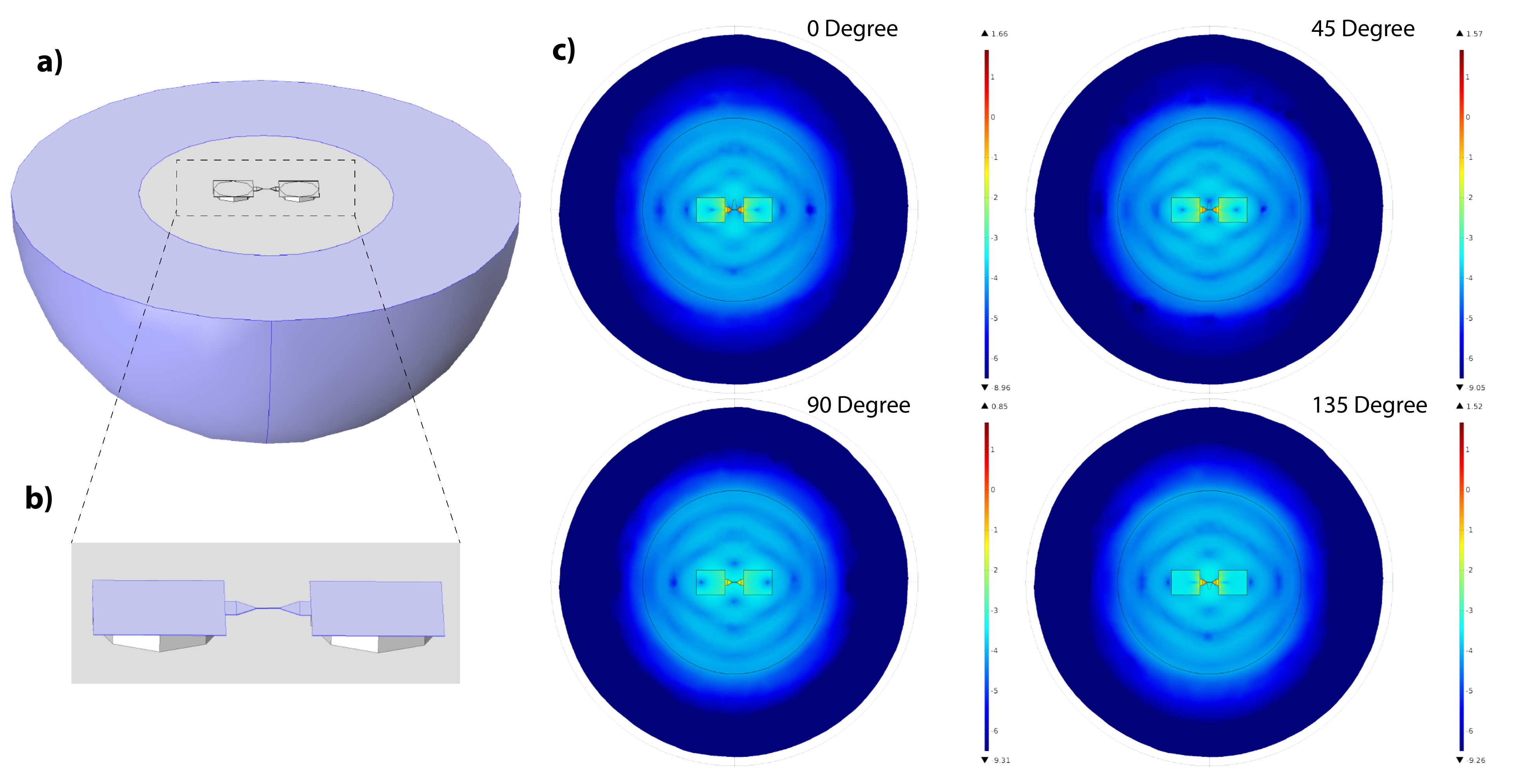}
		\caption{\textbf{FEM simulation of acoustic radiation loss.} (a) Geometry of the simulated device. The gray volume indicates the mechanical structure and the blue volume is a PML. (b) Magnified image of the mechanical structure, including the SiN beam and PnC shield, SiN support pads, Si pillars beneath the support pads, and a portion of the underlying Si substrate.  A stress relaxation simulation is performed in order to determine the stress profile of this structure before solving for eigenmodes. (c) Visualization of acoustic waves propagating into the Si substrate and being absorbed by PML.  The amplitude of the mechanical strain field is illustrated by logarithmic color-coding.}
		\label{fig:radiation_FEM}
		%\vspace{-3pt}
	\end{figure*}
	
	The radiation loss model shown in Fig. 4 of the main text was obtained by computing the complex eigenfrequency spectrum of the mechanical structure surrounded by numerically implemented PML (perfectly matched layer, corresponding to a perfectly impedance matched and aborbing boundary). We used the COMSOL Structural Mechanics Finite Element Analysis software package for this simulation.  The modeled structure is shown in gray in Fig. 2a, and includes the SiN defect beam, SiN PnC, SiN support pads, the Si pillars beneath each support pad, a half-spherical section of the Si substrate. The blue shell in Fig. 2a corresponds to the PML.  By placing the PML far from the nanobeam, we ensure that reflection of acoustic waves from the supports is accounted for in the simulation.  We also ensure that the size of the PML is large enough that its own reflection coefficient is negligible \cite{wilson2008intrinsic,wilson2011high}. Visualization of acoustic radiation into the PML is shown in Fig. 2c.  Radiation loss manifests as an imaginary component of each numerically computed eigenfrequency, $\Omega_\t{m}$.  The eigenfrequency of the defect mode is identified based on its mode shape.  Radiation-loss-limited $Q$-factors (blue curve in Fig. 4 of the main text) are obtained from the formula:
	\begin{equation}\label{eq:Qradversuscellnumber}
	Q_\text{rad}=\frac{\text{Re}[\Omega_\t{m}]}{2\cdot \text{Im}[\Omega_\t{m}]}.
	\end{equation}
	
	The location of the knee point in the model of $Q_\t{rad}$ versus $f$ in Fig. 4 of the main text depends on the dimensions of the extended mechanical structure, in particular that of Si pillars and the trapezoidal end-points of the beam (c.f. Fig. 1a in the main text).  The blue curve shown in Fig. 4 of the main text employs dimensions obtained from SEM imaging and uses no fit parameters.
	
	\subsection{B.\hspace{5mm}Role of number of unit cells}\label{sec:n_cell}
\begin{figure}[t]
		\centering
		\includegraphics[width=0.7\linewidth]{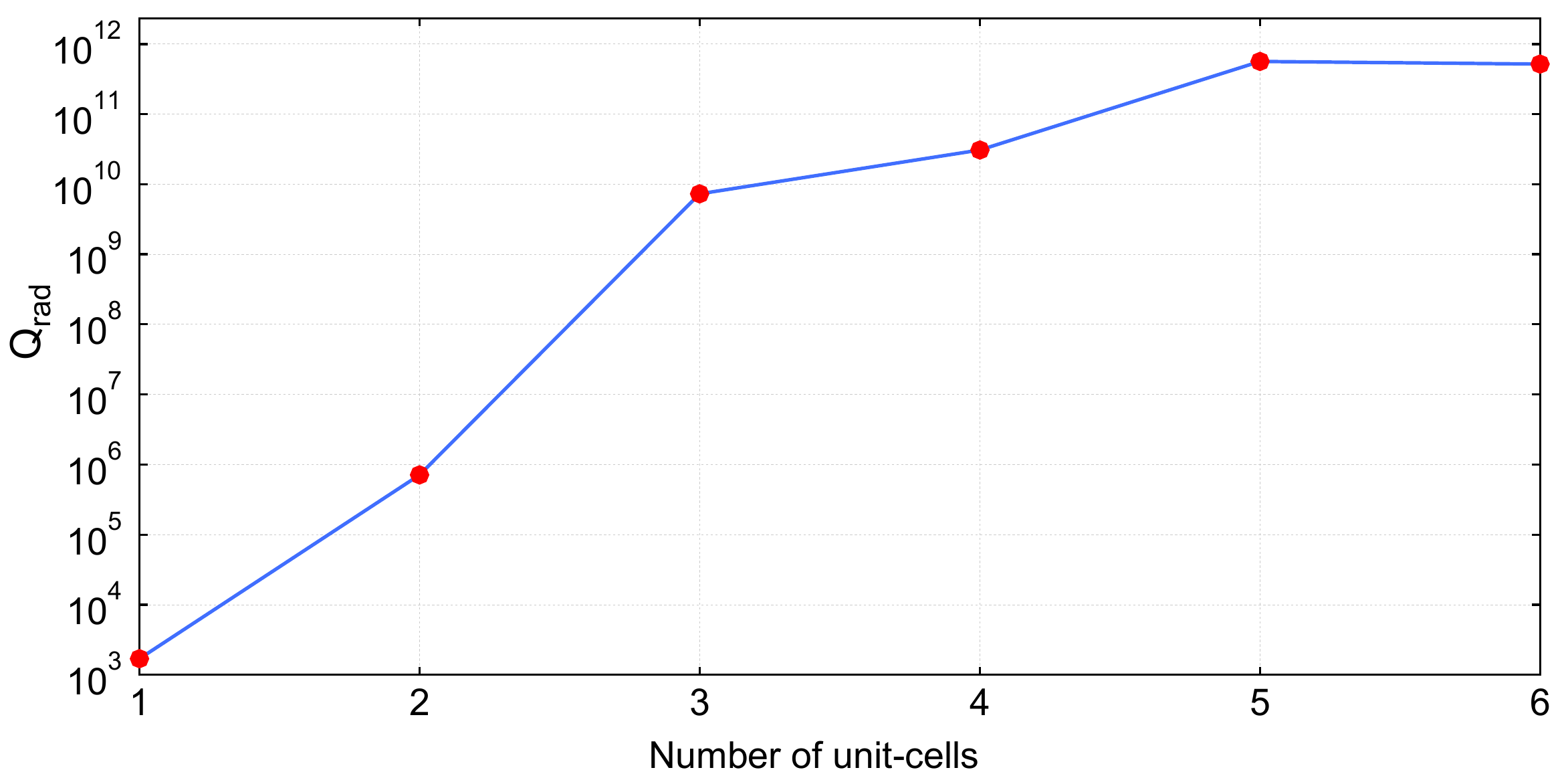}
		\caption{\textbf{Numerical simulation of acoustic radiation loss versus number of unit cells}. Results shown are for a 2.3 $\mu$m-long defect beam with a frequency of 542 MHz. The cell dimensions and band diagram are the same as in Fig. 2 of the main text.   }
		\label{fig:Q_n_cell}
		\vspace{-1.5 mm}
	\end{figure}

	The acoustic impedance of the PnC depends on the number of unit cells used.  We employed a 7-cell PnC to obtain the results described in the main text.  This choice was based on a model of $Q_\t{rad}$ versus number of cells.  An example is shown in \fref{fig:Q_n_cell}:  here a 2.3 $\mu$m defect with a frequency of 542 MHz is embedded in the 3rd bandgap of the PnC described in Fig. 2 of the main text.  As illustrated in the figure, only 3 cells are required to reduce the radiation loss by a factor of $10^6$, to a value $\sim6$ orders of magnitude larger than $Q_\t{int}$; also shielding evidently saturates at $\sim6$ cells.  A conservative choice of 7 cells was made in the experiment to allow for design imperfections.
	
	%\subsection{Radiation knee point}\label{sec:n_cell}
	%
	%The location of the knee point in the model of $Q_\t{rad}$ versus $f$ in Fig. 4 of the main text depends on the dimensions of the extended mechanical structure, in particular that of Si pillars and the trapezoidal end-points of the beam (c.f. Fig. 1a in the main text).  The blue curve shown in Fig. 4 of the main text employs dimensions obtained from SEM imaging and uses no fit parameters.  To illustrate the sensitivity of the model to device dimensions, in Fig. 4 we generate similar plots as a function of XXX...
	
	\vfill \pagebreak
%	\bibliographystyle{apsrev4-1}
%	\bibliography{ref1}

\end{document}